\documentclass[11pt,a4paper]{article}
\usepackage{jheppub}
\usepackage{amsfonts,amsmath,url}
\usepackage{eurosym}
\usepackage{braket}
\usepackage{graphicx}
\usepackage[utf8]{inputenc}
\usepackage{BOONDOX-cal}
\usepackage{xcolor}
\usepackage{ulem} 
\DeclareFontFamily{U}{MnSymbolC}{}
\DeclareSymbolFont{MnSyC}{U}{MnSymbolC}{m}{n}
\DeclareFontShape{U}{MnSymbolC}{m}{n}{
    <-6>  MnSymbolC5
   <6-7>  MnSymbolC6
   <7-8>  MnSymbolC7
   <8-9>  MnSymbolC8
   <9-10> MnSymbolC9
  <10-12> MnSymbolC10
  <12->   MnSymbolC12}{}
\DeclareMathSymbol{\intprod}{\mathbin}{MnSyC}{'270}

\begin{document}

\title{A tale of two theories of gravity in asymptotically Anti-de Sitter spacetime}


\author[a]{Remigiusz Durka}
\emailAdd{remigiusz.durka@uwr.edu.pl}
\author[a,b]{Jerzy Kowalski-Glikman}
\affiliation[a]{University of Wroc\l{}aw, Faculty of Physics and Astronomy, pl.\ M.\ Borna 9, 50-204 Wroc\l{}aw, Poland}
\affiliation[b]{National Centre for Nuclear Research, Pasteura 7, 02-093 Warsaw, Poland}
\emailAdd{jerzy.kowalski-glikman@uwr.edu.pl}

\abstract{We consider two BF formulations of the theory of gravity with a negative cosmological constant, of Plebanski and of MacDowell--Mansouri. Both give the standard Einstein equations in the bulk but differ in expressions of edge charges. We compute the asymptotic charges explicitly in both theories for AdS--Schwarzschild, AdS--Kerr, and AdS--Taub--NUT solutions. We find that while in the case of the Plebanski theory the charges are divergent, they are finite for MacDowell--Mansouri theory. Furthermore, we show that in the case of the arbitrary asymptotically AdS spacetimes, MacDowell--Mansouri asymptotic charges, action, and symplectic form are all finite. Therefore MacDowell--Mansouri theory of gravity in asymptotically AdS spaces does not need any counterterms.}

\maketitle

\section{Introduction}

In recent years we have witnessed an important shift in the front line of gravity research, both classical and quantum. Rather than investigate the bulk structure of gravity, described by solutions of Einstein (Wheeler--De Witt) equations, motivated by the AdS/CFT correspondence and discoveries of remarkable relations between asymptotic symmetries and soft theorems more and more interest is attracted by boundaries, finite asymptotic of spacetime regions.

These boundaries are commonly called corners and it was claimed that the Noether charges \cite{Kijowski:1976ze, Lee:1990nz, Iyer:1994ys, Brown:1992br, Crnkovic:1987tz, Wald:1999wa, Freidel:2015gpa, Frodden:2017qwh, Frodden:2019ylc, DePaoli:2018erh, Oliveri:2019gvm} associated with them, and their representations might carry relevant information, essential for the program of quantizing gravity \cite{Donnelly:2016auv}, \cite{Donnelly:2020xgu}.

In the recent series of papers \cite{Freidel:2020xyx, Freidel:2020svx, Freidel:2020ayo}, using the covariant phase space technique \cite{Crnkovic:1986ex, Crnkovic:1987tz, Julia:2002df, Harlow:2019yfa}, the authors analyze the corner charges structure in the case of two formulations of gravity, the traditional metric one and the Plebanski tetrad formulation expressed as a BF theory \cite{Plebanski:1977zz}. Following these investigations, in our paper \cite{Durka:2021ftc} we discussed the corner charges of gravity in the case of yet another formulation of gravity known as a constrained BF theory, proposed in \cite{Smolin:2003qu}, \cite{Freidel:2005ak} and based on the MacDowell and Mansouri approach \cite{MacDowell:1977jt}. These two theories give rise to the same bulk Einstein equations, but their Lagrangians differ by topological terms, which makes the corner charges very different. It is the aim of this paper to investigate these differences in detail, in the case of asymptotically Anti-de Sitter (AAdS) spacetime.

In the following section, we start our investigations of these two theories by writing their actions in the case of a non-vanishing negative cosmological constant. To this end the addition of the cosmological term is needed in the case of Plebanski theory, whereas the MacDowell--Mansouri model has a cosmological constant built-in. We normalize both theories so that their Lagrangians, up to the topological terms, agree with the Holst Lagrangian.

In Section 3 we recall the derivation of conserved corner charges in both theories. In the following Section 4, we use these obtained expressions to compute the asymptotic charges of AdS--Schwarzschild, AdS--Kerr, and AdS--Taub--NUT solutions. We find that these charges diverge for Plebanski theory, but they are finite in the case of  MacDowell--Mansouri model. 

In Section 5 we further investigate the MacDowell--Mansouri theory and we show that for the general asymptotic behavior of the metric all the charges of this theory are finite. We then show that the MacDowell--Mansouri action is finite. It follows that the symplectic structure is finite as well. This shows that, contrary to the Plebanski one, this theory does not require any counterterms at infinity \cite{Compere:2008us, Anastasiou:2020zwc}.

We finish the paper with short conclusions.


\section{The two BF theories}

In this section, we define two BF theories of gravity (for the extended discussion see \cite{Freidel:2012np}) in the case of spacetime with a negative cosmological constant and point out their similarities and differences. To distinguish them, we shall call the one considered in \cite{Freidel:2020xyx,Freidel:2020svx,Freidel:2020ayo} `Plebanski BF theory', while the one discussed in \cite{Durka:2021ftc} will be dubbed `MacDowell--Mansouri BF theory'.


\subsection{Plebanski BF theory}

The starting point here is the BF Lagrangian with the so-called 'cosmological term'
\begin{equation}\label{PlebBF}
 32\pi G\, \mathcal{L}_{Pl} =a\, B_{ij}\wedge R^{ij}(\omega) -\frac{b}2\, B_{ij}\wedge B^{ij}\,.
\end{equation}
In this Lagrangian $B_{ij}$ is a two-form valued in the Lorentz algebra $\mathsf{so}(3,1)$, $R^{ij}(\omega)$ is the curvature of Lorentz connection $\omega^{ij}$
\begin{equation}
    R^{ij}(\omega) =d \omega^{ij}+\omega^i{}_k\wedge \omega^{kj}\,,
\end{equation} 
and $a$, $b$ are parameters to be adjusted in a moment. The indices $i,j, \ldots=0,\ldots,3$ are lowered and raised with the help of the Minkowski metric $\eta_{ij}$. In the next step of the construction, we impose the simplicity constraint, which expresses the two form $B$ in terms of the tetrads, to wit
\begin{equation}\label{simplicity}
    B^{ij} =c\, e^i\wedge e^j + \epsilon^{ij}{}_{kl} e^k\wedge e^l\,.
\end{equation}
Substituting \eqref{simplicity} to \eqref{PlebBF} we obtain
\begin{equation}\label{Pleb}
 32\pi G\,  \mathcal{L}_{Pl} =a\,\epsilon_{ijkl}\, e^i\wedge e^j \wedge R^{kl} + ac\,e^i\wedge e^j  \wedge R_{ij} +{bc}\, \epsilon_{ijkl} e^i\wedge e^j \wedge e^k\wedge e^l \,.
\end{equation}
Comparing this with Holst Lagrangian \cite{Holst:1995pc},
\begin{equation}\label{Holst}
 32\pi G\,\mathcal{L}_{Holst} =\epsilon_{ijkl}\, e^i\wedge e^j \wedge R^{kl} + \frac2\gamma\,e^i\wedge e^j  \wedge R_{ij} -\frac{\Lambda}6\, \epsilon_{ijkl} e^i\wedge e^j \wedge e^k\wedge e^l \,,
\end{equation}
we see that $a =1$, $b=\Lambda\gamma/12$, $c=2/\gamma$. The second term in the action, known as the Holst term, is not topological but, as a result of vanishing torsion, it does not contribute to the bulk Einstein vacuum field equations. It influences, however, the symplectic form and, in turn, the form and algebra of (asymptotic) conserved charges.

The variation of Plebanski action is
\begin{equation}\label{action_Plebanski}
 32\pi G    \delta S_{Pl} = \delta \int B_{ij}\wedge R^{ij}(\omega) -\frac{\Lambda\gamma}{24}\, B_{ij}\wedge B^{ij}\,.
\end{equation}
Inserting the simplicity constraint \eqref{simplicity} we see that the $\delta e$ field equations are the standard Einstein equations, while the equations resulting from the variation of Lorentz connection $\delta \omega$ become the torsion vanishing condition. In the course of deriving this last equation, we have to integrate by parts, so that there is a boundary contribution to the variation of the action
\begin{equation}\label{varPleb}
 32\pi G    \delta S_{Pl} \approx   \int_\Sigma B_{ij}\wedge \delta \omega^{ij} =  \int_\Sigma \left(\frac2\gamma\, e^i\wedge e^j + \epsilon^{ij}{}_{kl} e^k\wedge e^l\right)\wedge \delta\omega_{ij}\,,
\end{equation}
where the weak equality $\approx$ denotes equality on-shell.
As explained in details in \cite{Freidel:2020svx} the resulting symplectic form further decomposes into the bulk $\Sigma$ and the corner $S=\partial\Sigma$ contributions.


\subsection{MacDowell--Mansouri BF theory}

In the case of MacDowell--Mansouri BF theory, the starting point is the BF Lagrangian with the cosmological term for the Anti-de Sitter gauge group appended by an additional deformation term that breaks the gauge symmetry down to the local Lorentz one\footnote{The more general case when the last term in the action depends on the background field ${\mathcal v}^I$ was considered in \cite{Durka:2021ftc}. It turns out that such generalization does not lead to any new effects, and therefore, without any loss of generality we restrict ourselves here to the case ${\mathcal v}^I=(0,0,0,0,-1)$.}
\begin{equation}\label{MMbBF}
    32\pi G\, \mathcal{L}_{MM} = B_{IJ}\wedge F^{IJ}(A) -\frac{\beta}2\, B_{IJ}\wedge B^{IJ} - \frac\alpha4\, \epsilon_{ijkl}\, B^{ij}\wedge B^{kl}\,,
\end{equation}
where $F^{IJ}(A)$ is the curvature of AdS connection $A^{IJ}$ and the indices $I,J, \ldots$ run from 0 to 4, $I=(i,4)$. The Anti-de Sitter connection $A^{IJ}$ decomposes into the Lorentz connection and the tetrad
\begin{equation}\label{AdScon}
   A^{ij} = \omega^{ij}\,,\quad A^{i4} = \frac1\ell\, e^i 
\end{equation}
with the parameter $\ell$ introduced for dimensional reasons to be given an interpretation in a moment.
The $B$-field equations can be decomposed to give
\begin{align}
B^{ij}&= \frac{\alpha}{(\alpha^{2}+\beta^{2})}\left(\frac{1}{2}\frac{\beta}{\alpha}\,\delta^{ij}_{kl}- \frac{1}{2}\epsilon^{ij}{}_{kl}\right)F^{kl}\,,\label{Bij}\\
B^{i4} &=  \frac{1}{\beta}\,F^{i4}\,,
\end{align}
where $\delta^{i j}_{k l}=\delta^{i}_{k}\delta^{j}_{l}-\delta^{i}_{l}\delta^{j}_{k}$ and the curvature $F^{IJ}=d A^{IJ}+A^{I}{}_{K}\wedge A^{KJ}$ components are
\begin{align}
    F^{ij}&=R^{ij}(\omega )+\frac{1}{\ell ^{2}}\,e^{i}\wedge e^{j}\,,\\
    F^{i4}&=\frac{1}{\ell }T^{i}\,.
\end{align}
In what follows, we will call $F^{ij}$ the AdS-curvature, to distinguish it from the Lorentz curvature $R^{ij}(\omega)$. 
Substituting these equations to the original Lagrangian and using the explicit expressions for the parameters\footnote{In the formulas below, we make the change $\gamma\mapsto-1/\gamma$ as compared to the standard formulation of MM BF theory \cite{Freidel:2005ak, Durka:2012wd, Durka:2021ftc} to have the inverse Barbero-Immirzi parameter as a coefficient of Holst term, in agreement with most of the literature.}
\begin{align}\label{constant}
   \alpha= \frac{ \Lambda}{3}\frac{\gamma^2}{1+{\gamma^2}}\,,\qquad {\beta}=-\frac{\alpha}\gamma\,,\qquad \Lambda=- \frac{3}{\ell^2} \,,
\end{align}
we obtain
\begin{align}
 32\pi G\,\mathcal{L}_{MM}& = \epsilon_{ijkl}R^{ij}\wedge e^{k}\wedge \,e^{l}-\frac{\Lambda}{6}\epsilon_{ijkl} e^{i}\,\wedge e^{j}\wedge e^{k}\,\wedge
e^{l}  -{2}{\gamma }R_{ij}\wedge e^{i}\wedge e^{j} \nonumber \\
& -2 \frac{1+\gamma^2}{\gamma }\,\left(T^{i}\wedge T_{i}-R_{ij}\wedge e^{i}\wedge
e^{j}\right) -\frac{3}{\Lambda\gamma}R^{ij}\wedge R_{ij}-\frac{3}{2\Lambda}\epsilon_{ijkl}R^{ij}\wedge R^{kl}\,.\label{actionA.11}
\end{align}
The first line in $\mathcal{L}_{MM}$ reproduces the Holst Lagrangian \eqref{Holst}, albeit with the inverse power of the Barbero-Immirzi parameter, while the second line is a combination of Nieh-Yan, Pontryagin, and Euler forms. One then combines the second term of the Nieh-Yan form with the Holst term in the first line to finally get the right $2/\gamma$ coefficient.

The Lagrangian \eqref{actionA.11} can be conveniently rewritten as a sum of squares of torsion and AdS curvature
\begin{align}\label{actionA.11b}
 32\pi G\,\mathcal{L}_{MM}
&=-\frac{3}{2\Lambda}\left(\frac{1}{\gamma}\delta_{ijkl}+\epsilon_{ijkl}\right)F^{ij}\wedge F^{kl}-2 \frac{1+\gamma^2}{\gamma }\,T^{i}\wedge T_{i}\,.
\end{align}

Let us pause here to comment on why the theories considered here can be, as it is often done, called ``constrained BF theories''. The Lagrangians of Plebanski BF theory with a cosmological constant \eqref{action_Plebanski} and the MacDowell--Mansouri BF theory \eqref{MMbBF} can be both written as a sum of the topological BF Lagrangian accompanied by the Lagrange multiplier enforcing an appropriate constraint
\begin{align}
 32\pi G\,\mathcal{L}_{Pl}&\approx B_{ij}\wedge R^{ij}-\frac{\Lambda\gamma}{24}\, B_{ij}\wedge B^{ij}+\lambda_{ij} \left(B^{ij}-\left(\frac{1}{\gamma}\delta^{ij}_{kl}+\epsilon^{ij}{}_{kl}\right) e^k\wedge e^l\right)\label{BFPL}\\  
 32\pi G\,\mathcal{L}_{MM}&\approx B_{IJ}\wedge F^{IJ}+ \lambda_{ij} \left(B^{ij}+\frac{3}{2\Lambda}\left(\frac{1}{\gamma}\delta^{ij}_{kl}+\epsilon^{ij}{}_{kl}\right)F^{kl}\right) + \lambda_i\left(B^{i4} - F^{i4}\right)\label{BFMM} \,. 
\end{align}
In spite of the fact that the theories are very different, it follows from \eqref{BFPL} and \eqref{BFMM} that to derive the symplectic form and conserved charges  resulting from the first BF term in the Lagrangian, one can handle both theories in an essentially the same way inserting (the same) bulk field equations and (different) fields $B$ at the very end of the calculations.

Now that we have described the formulation of the two BF theories, let us turn to investigate their further properties.


\section{The conserved charges}

In this section, we will describe the conserved surface charges associated with local gauge symmetries. Both BF theories described in the preceding section are invariant under local Lorentz transformations and diffeomorphisms, and there are conserved charges associated with these symmetries. These charges are defined by the formula
\begin{align}
    -\delta {\cal H}_s = \delta_s \intprod \Omega\,,
\end{align}
where  ${\cal H}_s$ is the charge associated with symmetry $\delta_s $ and $\Omega$ is the symplectic form, which has the form
\begin{align}
    \Omega_{Pl} &= \frac{1}{ 32\pi G}\int_\Sigma \delta B^{ij}\wedge \delta \omega_{ij} \label{omegaPl}\\
    \Omega_{MM} &= \frac{1}{ 32\pi G}\int_\Sigma \delta B^{IJ}\wedge \delta A_{IJ} \,.\label{omegaMM}
\end{align}
These symplectic forms can be further decomposed into the bulk and boundary parts, but since we will investigate the resulting charges in the case of the given on-shell asymptotic expansion of the gravitational field, we do not need to make use of this decomposition here.


\subsection{Plebanski theory }

Let us start with Lorentz symmetry with parameter $\lambda^{ij}$. Since
\begin{align}
    \delta \omega^{ij} = d\lambda^{ij} + [\omega, \lambda]^{ij}\,,\quad \delta B^{ij} =  [B, \lambda]^{ij}\,,
\end{align}
we have
\begin{align*}
  -  \delta {\cal H}_{Pl}[\lambda] = \frac{1}{ 32\pi G}\int_\Sigma [B, \lambda]^{ij} \wedge \delta \omega_{ij} - \delta B_{ij}\wedge \left(d\lambda^{ij} + [\omega, \lambda]^{ij}\right) =-\frac{1}{ 32\pi G} \int_S \delta\left( B_{ij}\lambda^{ij}\right)\,,
\end{align*}
so that the Lorentz charge has only the corner component
\begin{equation}
 {\cal H}_{Pl}[\lambda] =   \frac{1}{ 32\pi G} \int_S  B_{ij}\lambda^{ij} =  \frac{1}{ 32\pi G} \int_S \left(\frac2\gamma\, e^i\wedge e^j + \epsilon^{ij}{}_{kl} e^k\wedge e^l\right) \lambda_{ij}\,,\label{Hlambda}
\end{equation}
where to get the second equality we used the simplicity constraint.

Next, consider the charge associated with the diffeomorphisms. Following \cite{Freidel:2020xyx} we have 
\begin{equation}
\delta {\cal H}_{Pl}[\xi] =-\mathcal{L}_{\xi}\intprod \Omega = \frac{1}{ 32\pi G}\int_\Sigma \delta B_{ij}\wedge\mathcal{L}_{\xi}  \omega^{ij}- \mathcal{L}_{\xi} B_{ij}\wedge\delta \omega^{ij}\,,\label{deltaHxi}
\end{equation}
where $\mathcal{L}_{\xi}$ is a Lie derivative and we used an identity $\mathcal{L}_{\xi}(\cdot)=\xi\lrcorner (d\,  \cdot)+d(\xi\lrcorner\, \cdot)$. This can be further rewritten as
\begin{equation}
\delta {\cal H}_{Pl}[\xi] =-\mathcal{L}_{\xi}\intprod \, \Omega=\frac{1}{ 32\pi G}\delta\left( \int_{\Sigma} B_{ij} \wedge \mathcal{L}_{\xi} \omega^{ij}\right)-\frac{1}{ 32\pi G}\int_{S} \xi\intprod\left(B_{ij} \wedge \delta \omega^{ij}\right)\,.\label{deltaHxi1}
\end{equation}
If we assume that the only non-vanishing components of the vector field $\xi$ on the corner $S$ are those tangent to $S$, the second term is a diffeomorphism acting on an integral of a scalar density and vanishes identically. Therefore, in this case, the diffeomorphism charge has the form
\begin{equation}
\mathcal{H}_{Pl}[\xi]=\frac{1}{ 32\pi G}\int_{\Sigma} B_{ij} \wedge \mathcal{L}_{\xi} \omega^{ij}\,.\label{Hxi}
\end{equation}
However, for the diffeomorphism generated by an arbitrary vector field $\xi$, the last term in \eqref{deltaHxi1} does not vanish, and the charge ${\cal H}_\xi$ is not well defined. This case was discussed in detail in \cite{Freidel:2021cjp}. Here we resolve this problem by using the definition of the diffeomorphism charge proposed by Wald \cite{Wald:1993nt}, in which the definition \eqref{deltaHxi} is modified by the addition of the corner term that cancels the trouble-making one in \eqref{deltaHxi1}
\begin{equation}
    \delta {\cal H}_{Pl}^W[\xi] = \delta {\cal H}_\xi +\frac{1}{ 32\pi G} \int_{S} \xi\intprod\left(B_{ij} \wedge \delta \omega^{ij}\right)\,.
\end{equation}
In this way, \eqref{Hxi} becomes the expression for the diffeomorphism charge for arbitrary vector field $\xi$ that we are going to adopt here. In what follows, we will use the Wald definition for diffeomorphism charge, and we will not use the label $W$ anymore.

Using the identity $\mathcal{L}_{\xi} \omega^{ij} = \xi \lrcorner R(\omega)^{ij}+D^{\omega}(\xi \lrcorner \omega^{ij})$, which is a consequence of the field equations $D^\omega B_{ij}=0$, we write \eqref{Hxi} as
\begin{equation}
\mathcal{H}_{Pl}[\xi]=\frac{1}{ 32\pi G}\int_{\Sigma}  B_{ij} \wedge \xi \lrcorner R(\omega)^{ij}+d(B_{ij} \wedge  \xi \lrcorner \omega^{ij}) \,,
\end{equation}
which decomposes into the bulk part
\begin{equation}
   {\cal H}_{Pl}^\Sigma[\xi]  =\frac{1}{ 32\pi G}\int_{\Sigma}  B_{ij} \wedge \xi \lrcorner R(\omega)^{ij}
\end{equation}
and the corner component
\begin{equation}
\mathcal{H}_{Pl}^{S}[\xi]=\frac{1}{ 32\pi G} \int_{S}  B_{ij} \wedge  \xi \lrcorner \omega^{ij}\,.\label{H_corner_Pl}
\end{equation}
Using $B^{ij}=\left(\frac2\gamma\, e^i\wedge e^j + \epsilon^{ij}{}_{kl} e^k\wedge e^l\right)$ we can write latter as
\begin{equation}
\mathcal{H}_{Pl}^{S}[\xi]= \frac{1}{ 32\pi G}\int_{S} \left(\frac2\gamma\, e^i\wedge e^j + \epsilon^{ij}{}_{kl} e^k\wedge e^l\right)\wedge  \xi \lrcorner \omega^{ij}\,,\label{H_corner_Pl_explicit}
\end{equation}
which encapsulates the dual ('magnetic') and standard ('electric') parts of the charge \cite{Godazgar:2020kqd, Godazgar:2020gqd}.

One immediately sees the potential problem that could arise in the case of asymptotic charges of Plebanski theory \eqref{H_corner_MM_explicit}. The integrand of this charge goes to zero rather slowly as $r\rightarrow\infty$ and one may wonder if the asymptotic charges are finite. As we will see in the next section this is indeed the case, even for the simplest asymptotically AdS solutions of Einstein equations.


\subsection {MacDowell--Mansoury theory}

In the case of MacDowell--Mansouri BF theory, the derivation of the expressions for charges goes in exactly the same way as above, so we can readily write for the Lorentz charges
\begin{equation}
 {\cal H}_{MM}[\lambda] =  \frac{1}{ 32\pi G} \int_S  B_{ij}\lambda^{ij}  \label{HlambdaMM}
\end{equation}
 and for the diffeomorphisms ones
\begin{equation}
\mathcal{H}_{MM}[\xi]\thickapprox\frac{1}{ 32\pi G}\int_{\Sigma}  B_{IJ} \wedge \xi \lrcorner F(A)^{IJ}+d(B_{IJ} \wedge  \xi \lrcorner A^{IJ}) \,,
\end{equation}
which again can be decomposed into the bulk and corner parts. Using the fact that $B_{I4}=0$ from field equations (vanishing of torsion), we have 
\begin{equation}
   {\cal H}_{MM}^\Sigma[\xi]  =\frac{1}{ 32\pi G}\int_{\Sigma}  B_{ij} \wedge \xi \lrcorner F(A)^{ij}
\end{equation}
and the corner component
\begin{equation}
\mathcal{H}_{MM}^{S}[\xi]= \frac{1}{ 32\pi G}\int_{S}  B_{ij} \wedge  \xi \lrcorner \omega^{ij}\,.\label{H_corner_MM}
\end{equation}
In these equations $B_{ij}$ should be replaced by the on-shell value of the expression on the right-hand side of \eqref{Bij}, ie.,
\begin{align*}
   B^{ij}=-\frac{1}{2}\,\frac{3}{\Lambda}\left(\frac1\gamma\delta_{kl}^{ij}+\epsilon^{ij}{}_{kl}\right) F^{kl}\,,
\end{align*}
which for the corner charge gives
\begin{align}
\mathcal{H}_{MM}^{S}[\xi] &=- \frac{3}{ 64\pi\Lambda G}\int_{S}  \left(\frac1\gamma\delta_{kl}^{ij}+\epsilon^{ij}{}_{kl}\right)\,F^{kl}  \xi \lrcorner \omega_{ij}\,.\label{H_corner_MM_explicit}
\end{align}

Contrary to the Plebanski BF theory, here the integrand goes to zero fast asymptotically, which suggests that the asymptotic charges are finite. Below we will show that this is indeed the case.


\section{The charges of AdS--Schwarzschild, AdS--Kerr, and AdS--Taub--NUT}

Before discussing the most general case of the arbitrary asymptotically AdS spacetime, as a warm-up, let us compute the asymptotic charges of two simple solutions, AdS--Schwarzschild and AdS--Kerr, whose physical interpretation is clear. Then we consider the more complicated AdS--Taub--NUT. As we will see, the charges of MacDowell--Mansouri BF theory have the expected finite form, while the ones of Plebanski BF theory are divergent. This suggests that Plebanski BF theory requires   addition of the counterterms that render the charges finite, see e.g. \cite{Compere:2008us}, \cite{Anastasiou:2020zwc}, and references therein.


\subsection{AdS--Schwarzschild}

The AdS--Schwarzschild metric reads
\begin{align}
    ds^2 = - \left(1-\frac{2 G M}{r}- \frac{\Lambda}{3}r^{2}\right) dt^2 + \left(1-\frac{2 G M}{r}- \frac{\Lambda}{3}r^{2}\right)^{-1} dr^2 + r^2 d\Omega^2\,.
\end{align}
The corresponding tetrads are defined as
\begin{align}
  e^0=f(r) dt,\qquad e^1=f(r)^{-1}dr,\qquad e^2=r d\theta, \qquad e^3=r \sin\theta d\varphi
\end{align}
with
\begin{equation}
    f(r)^{2}=1-\frac{2 G M}{r}- \frac{\Lambda}{3}r^{2}\,,
\end{equation}
and one derives expressions for the connection $\omega^{ij}$ 
\begin{align}
\omega^{01} &=f'(r) e^{0} &\omega^{23} &=-\frac{1}{r} \frac{\cos\theta}{\sin\theta} e^{3}\nonumber\\
\omega^{12} &=-\frac{f(r)}{r} e^{2} &\omega^{13} &=-\frac{f(r)}{r} e^{3}\label{AdSS-connection}\,.
\end{align}
The AdS curvature $F^{ij}=R^{ij}+\frac{1}{\ell^2}e^i \wedge e^j$ of the AdS--Schwarzschild spacetime can be written in a simple form as
\begin{align}
F^{01}&=-2 h e^{0} \wedge e^{1} &F^{02}&=h e^{0} \wedge e^{2} \nonumber\\
F^{03}&=h e^{0} \wedge e^{3} &F^{12}&=h e^{1} \wedge e^{2}\nonumber\\
F^{13}&=h e^{1} \wedge e^{3} &F^{23}&=-2 h e^{2} \wedge e^{3}\label{AdSS-curvature}
\end{align}
with the $h=-{G M}/{r^3}$. Notice that the AdS-curvature goes to zero asymptotically.

For the calculation concerning mass, we take $\partial_t$ Killing vector, which naturally leads to the asymptotic corner diffeomorphism charge, computed on the sphere at infinity $S_\infty$ 
\begin{align}
\mathcal{H}_{MM}^{S_\infty}[\partial_t]&    =M\,,
\end{align}
providing the expected value of the charge at infinity. The Plebanski theory gives instead the divergent result
\begin{align}
\mathcal{H}_{Pl}^{S_\infty}[\partial_t]&=\frac{M}{2}-\lim_{r\to\infty} \frac{\Lambda }{3}\frac{r^{3}}{2G}
\end{align}
with the finite part now being half of the mass parameter.
This shows that Plebanski theory requires holographic renormalization, which not only renders the asymptotic charges finite but also rescales the finite part of the charge. It can be checked that adding the Euler term to Plebanski action, with the coefficient $-3/2\Lambda$ being exactly that of MacDowell--Mansouri theory cf.\ \eqref{actionA.11}, does the trick here \cite{Aros:1999id, Aros:1999kt}.

It can be further checked, that all other diffeomorphism charges vanish for both theories.


\subsection{AdS--Kerr}

Tetrads of the AdS--Kerr are given by
\begin{align}
e^{0} &=\frac{\sqrt{\Delta_{r}}}{\rho}\left(d t-\frac{a}{\Xi} \sin ^{2} \theta d \varphi\right), \quad e^{1}=\rho \frac{d r}{\sqrt{\Delta_{r}}} \\
e^{2} &=\rho \frac{d \theta}{\sqrt{\Delta_{\theta}}}, \quad e^{3}=\frac{\sqrt{\Delta_{\theta}}}{\rho} \sin \theta\left(\frac{\left(r^{2}+a^{2}\right)}{\Xi} d \varphi-a d t\right)\,,
\end{align}
where we use the standard abbreviations
\begin{align}
\rho^{2}&=r^{2}+a^{2} \cos ^{2} \theta\,,\\ \Delta_{r}&=\left(r^{2}+a^{2}\right)\left(1-\frac{\Lambda r^{2}}{3}\right)-2 G M r\,, \\
\Delta_{\theta}&=1+\frac{\Lambda a^2}{3} \cos^{2} \theta\,,\\ 
\Xi&=1+\frac{\Lambda a^2}{3}\,.
\end{align}
The volume element reads
\begin{align*}
    e^{0}\wedge  e^{1}\wedge  e^{2}\wedge  e^{3}=\frac{r^2+a^2\cos^2\theta}{\Xi}dt\wedge dr\wedge d\theta\wedge d\varphi\,.
\end{align*}

The non-vanishing connection components read
$$
\begin{aligned}
\omega^{01}&=\left(\frac{(r^{2}+a^{2} \cos ^{2} \theta)( M-r+(2r^2+a^2)\frac{\Lambda}{3}r)+r\Delta_r)}{\sqrt{\Delta_r}\rho^{3/2}}\right)e^0-\frac{a r \sqrt{\Delta_{\theta}} \sin \theta}{\rho^{3 / 2}} e^{3} \\
\omega^{02} &=-\frac{a^{2} \sqrt{\Delta_{\theta}} \cos \theta \sin \theta}{\rho^{3 / 2}} e^{0}-\frac{a \sqrt{\Delta_r} \cos \theta}{\rho^{3 / 2}} e^{3} \\
\omega^{03} &=-\frac{a r \sqrt{\Delta_{\theta}} \sin \theta}{\rho^{3 / 2}} e^{1}+\frac{a \sqrt{\Delta_r} \cos \theta}{\rho^{3 / 2}} e^{2} \\
\omega^{12} &=-\frac{a^{2} \sqrt{\Delta_{\theta}} \cos \theta \sin \theta}{\rho^{3 / 2}} e^{1}-\frac{r \sqrt{\Delta_r}}{\rho^{3 / 2}} e^{2} \\
\omega^{13} &=-\frac{a r \sqrt{\Delta_{\theta}} \sin \theta}{\rho^{3 / 2}} e^{0}-\frac{r \sqrt{\Delta_r}}{\rho^{3 / 2}} e^{3} \\
\omega^{23} &=-\frac{a \sqrt{\Delta_r} \cos \theta}{\rho^{3 / 2}} e^{0}-\left(\frac{\left(a^{2}+r^{2}\right)+a^{2} \frac{\Lambda}{3}\left(a^{2}+r^{2}-2\left(a^{2}+r^{2}\right) \sin^2\theta+a^{2} \sin^4 \theta\right)}{\rho^{3 / 2} \sqrt{\Delta_{\theta}}}\right) \frac{\cos\theta}{\sin\theta} e^{3}
\end{aligned}
$$
Quite unexpectedly, the AdS curvatures $F^{ij}=R^{ij}-\frac{\Lambda}{3}e^i \wedge e^j$ components have a simple form \cite{Blagojevic:2020edq, Blagojevic:2020ymf}
\begin{align}
F^{01}&=-2 h e^{0} \wedge e^{1}-2 b e^{2} \wedge e^{3} &F^{02}&=h e^{0} \wedge e^{2}-b e^{1} \wedge e^{3} \nonumber\\
F^{03}&=h e^{0} \wedge e^{3}+b e^{1} \wedge e^{2} &F^{12}&=h e^{1} \wedge e^{2}-b e^{0} \wedge e^{3} \nonumber\\
F^{13}&=h e^{1} \wedge e^{3}+b e^{0} \wedge e^{2} &F^{23}&=-2 h e^{2} \wedge e^{3}+2 b e^{0} \wedge e^{1}
\end{align}
with the  functions $h$, $b$ computed to be
\begin{align}
h&=\frac{\left(3 a^{2} \cos^2\theta-r^{2}\right)r}{\left(r^{2}+a^{2} \cos^2\theta\right)^{3}} M \label{h_Kerr}\\
b&=\frac{\left(3 r^{2}-a^{2} \cos^2\theta\right) a \cos\theta}{\left(r^{2}+a^{2} \cos^2\theta\right)^{3}} M \label{b_Kerr}\,.
\end{align}

We focus on the asymptotic Killing vectors of time translation $ \partial_t$ and rotation around the $z$-axis of axial symmetry $ \partial_\varphi$. In the case of the MacDowell--Mansouri theory, we obtain the following expressions for the mass and angular momentum 
\begin{align}
\mathcal{H}^{S}_{MM}[\partial_{t}]&=\frac{M}{\Xi } \\
\mathcal{H}_{MM}^{S}[\partial_\varphi] &=\frac{-aM}{\Xi ^{2}}\,.
\end{align}
It can be checked that the corner charges associated with other asymptotic Killing vectors vanish.

There is an ambiguity discussed, for example in \cite{Gibbons:2004ai, Durka:2012wd} concerning the proper definition of the mass. Taking into account the presence of an non-zero angular velocity at infinity, instead of the Killing vector $\partial_t$ it is more proper to use $\tilde\partial_t=\partial_t+\frac{a \Lambda}{3}\partial_\varphi$, which leads to the modified expression for the mass
\begin{align}\label{proper_Kerr_mass}
    \mathcal{H}_{MM}^{S}[\tilde\partial_t]= \frac{M}{\Xi^2}\,.
\end{align}

In the case of the Palatini BF theory, for the asymptotic charge associated with time translation we find 
\begin{align}
\mathcal{H}_{Pl}^{S}[\partial_{t}]&=\lim_{r\to\infty} \left( \frac{M/2%
}{\Xi }-\frac{\Lambda }{3}\frac{1}{2G}\frac{(r^{2}+a^{2})r}{\Xi }\right) =
\frac{M}{2\Xi }-\lim_{r\to\infty}  \frac{\Lambda }{6G}\frac{r^{3}}{\Xi }\,.
\end{align}
We see again that the mass is divergent and the finite part is half of the expected value, as it was in the Schwarzschild case. Exactly as it was in the Schwarzschild case, adding the Euler counterterm with the coefficient of MM theory to the Plebanski Lagrangian renders the finite part correct and cancels the divergent term.

As for the angular momentum, the result agrees with the one for MacDowell--Mansouri theory
\begin{align}
\mathcal{H}_{Pl}^{S}[\partial_{\varphi }] =\frac{-aM}{\Xi ^{2}}\,.
\end{align}

Interestingly, in the Plebanski BF theory, in the leading and subleading $r$ order, we have a non-vanishing, divergent charges of the boost $\mathcal{N}_3$ and translation $\mathcal{P}_3$ along the $z$ axis
\begin{align}
\mathcal{H}_{Pl}^{S}[\mathcal{N}_3] & =- \frac{1}{\gamma}\sin \left(\sqrt{-\Lambda/3 }\, t\right) \lim_{r\to\infty} \frac{\frac{\Lambda}{3}a(r^2+ a^2) }{6G \Xi}
\\
\mathcal{H}_{Pl}^{S}[\mathcal{P}_3] &= -\frac{1}{\gamma} \cos \left(\sqrt{-\Lambda/3 }\, t\right)  \lim_{r\to\infty} \frac{\left(-\frac{\Lambda}{3}\right)^{3/2}a(r^2+ a^2)}{6G \Xi}\,.
\end{align}
These expressions can be again canceled by adding the topological term.
All the other asymptotic charges vanish.


\subsection{AdS--Taub--NUT solution}

In both cases discussed above, the asymptotic charges did not depend on the Barbero-Immirzi parameter. This changes in the case of AdS--Taub--NUT solution, as we are going to show now. This solution attracted some attention recently in the context of holography; see for example \cite{Kalamakis:2020aaj} and references therein. In what follows we do not impose any conditions on the AdS-Taub-NUT parameters, contrary, for example, to  the Euclidean approach of \cite{Liko:2011cq, Araneda:2016iiy, Ciambelli:2020qny, Corral:2021xsu}.

The AdS--Taub--NUT spacetime is defined by the tetrad one forms
\begin{align}
  & e^0=f(r) (dt-2n(\cos\theta -k) d\varphi),\nonumber\\ &e^1=f(r)^{-1}dr,\nonumber\\
  &  e^2=\sqrt{r^2+n^2} d\theta,\nonumber\\ & e^3=\sqrt{r^2+n^2} \sin\theta d\varphi
\end{align}
with
\begin{equation}
    f(r)^{2}=\frac{r^{2}-2 G M r-n^{2}-\Lambda\left(r^{4}+6 n^{2} r^{2}-3 n^{4}\right)/3}{n^{2}+r^{2}}\,.
\end{equation}
Parameter $k=\{-1,0,1\}$ encodes location of the so-called Misner string(s) \cite{Misner:1963fr} with $k=0$ representing case with strings on both poles, whereas $k=-1$ one string up, and $k=1$ one string down.

The spin connection components are given by
\begin{align}
\omega^{01} &=f'(r) e^{0}\nonumber\\
\omega^{02} &=\frac{n f(r)}{n^{2}+r^{2}} e^{3} \nonumber\\
\omega^{03} &=-\frac{n f(r)}{n^{2}+r^{2}} e^{2} \nonumber\\
\omega^{12} &=-\frac{r f(r)}{n^{2}+r^{2}} e^{2}\nonumber \\
\omega^{13} &=-\frac{r f(r)}{n^{2}+r^{2}} e^{3} \nonumber\\
\omega^{23} &=\frac{n f(r)}{n^{2}+r^{2}} e^{0}-\frac{1}{\sqrt{n^{2}+r^{2}}} \frac{\cos\theta}{\sin\theta} e^3\,, 
\end{align}
whereas AdS-curvatures are 
\begin{align}\label{AdS-spanned}
F^{01}&=-2 h e^{0} \wedge e^{1}-2 b e^{2} \wedge e^{3} &F^{02}&=h e^{0} \wedge e^{2}-b e^{1} \wedge e^{3} \nonumber\\
F^{03}&=h e^{0} \wedge e^{3}+b e^{1} \wedge e^{2} &F^{12}&=h e^{1} \wedge e^{2}-b e^{0} \wedge e^{3} \nonumber\\
F^{13}&=h e^{1} \wedge e^{3}+b e^{0} \wedge e^{2} &F^{23}&=-2 h e^{2} \wedge e^{3}+2 b e^{0} \wedge e^{1}\,,
\end{align}
where $h$ and $b$ are two auxiliary functions
\begin{align}
h&=\frac{\left(3 n^{2}-r^{2}\right) r}{\left(n^{2}+r^{2}\right)^{3}} \,G M+\frac{\left(n^{2}-3 r^{2}\right) n}{\left(n^{2}+r^{2}\right)^{3}}\left(1-\frac{4 \Lambda n^{2}}{3}\right) n \,,\label{TNUTh}\\
b&=-\frac{\left(3 n^{2}-r^{2}\right) r}{\left(n^{2}+r^{2}\right)^{3}}\left(1-\frac{4 \Lambda n^{2}}{3}\right) n+\frac{\left(n^{2}-3 r^{2}\right) n}{\left(n^{2}+r^{2}\right)^{3}}\, G M\label{TNUTb} \,.
\end{align}
We notice that the AdS--Taub--NUT solution is defined by two parameters, the mass $M$ and the dual mass $N=n(1-4n^2 {\Lambda}/{3})/G$ \cite{Godazgar:2022jxm}. When $\Lambda \to 0$ we recover Taub--NUT solution, while in the limit $n\to0$ we get AdS--Schwarzschild spacetime with $b=0$ and $h=-{G M}/{r^3}$. Asymptotically, for $r\to \infty$ the leading order expansion is
\begin{equation}
    h \to -\frac{G M}{r^3} \,,\qquad b \to -\frac{N}{r^3}\,,\quad f\to \sqrt{\frac{|\Lambda|}3} r\,.
\end{equation}

As it was pointed out in \cite{Misner:1963fr, Zanelli:2009kra}, the Killing vectors are now modified
\begin{align}
    \begin{aligned}
\xi^{t} &=\frac{\partial}{\partial t} \\
\xi^{x} &=-\sin \varphi \frac{\partial}{\partial \theta}-\cos \varphi \cot \theta \frac{\partial}{\partial \varphi}-2 n k\left( \cos \varphi \cot \theta-\frac{\cos \varphi}{\sin \theta}\right) \frac{\partial}{\partial t} \\
\xi^{y} &=\cos \varphi \frac{\partial}{\partial \theta}-\sin \varphi \cot \theta \frac{\partial}{\partial \varphi}-2 n k\left( \sin \varphi \cot \theta-\frac{\sin \varphi}{\sin \theta}\right) \frac{\partial}{\partial t} \\
\xi^{z} &=\frac{\partial}{\partial \varphi}+2 n k \frac{\partial}{\partial t}\,.
\end{aligned}
\end{align}

The corner charge  computed for the time-like vector $\xi^t$ has contributions from the mass (resulting from Einstein-Cartan + Euler  \cite{Aros:1999kt, Aros:1999id}) and from the dual mass (appearing as a consequence of the Holst and Pontryagin terms \cite{Durka:2011yv})
\begin{align}
\mathcal{H}_{MM}^{S}[\xi^t]&=  M+\frac{1}{\gamma} \left(1-\frac{4 \Lambda  n^2}{3}\right)\frac{n}{G}\,.
\end{align}
The charge associated with the rotational Killing vector, somewhat similar to the AdS--Kerr case above, requires not taking $\partial_\phi$ but $\xi^{z}$. This gives the quantity associated with angular momentum
\begin{align}
\mathcal{H}_{MM}^{S}[\xi^z]&=  - k n \left(M +\frac1\gamma\,\left(1-\frac{4 \Lambda  n^2}{3}\right)\frac{n}{G}\right)\,.
\end{align}
It was stressed in \cite{Bonnor:1969ala, Manko:2005nm, Durka:2019ajz, Frodden:2021ces}, the Misner strings could be considered as the source of angular momentum. In the case of a single string (either up, $k=-1$ or down, $k=+1$) the angular momentum is proportional to $n\cdot (\mbox{Mass})$. This is analogous to the case of  Kerr metric, where for the Kerr parameter $a$, we have angular momentum proportional to $a \cdot (\mbox{Mass})$. Choosing $k=0$ leads to cancellation of the total angular momentum, as then we deal with two counter-rotating Misner stings with angular velocity being $\Omega=\pm \frac{1}{2n}$.

The corresponding charges calculated for the Plebanski theory are 
\begin{align}
\mathcal{H}_{Pl}^{S}[\xi^t]&= \frac1{\gamma}
\left(\frac{n}{2 G}-\frac{\Lambda  n \left(5 n^2+r^2\right)}{6 G}\right)-\frac{\Lambda  r \left(n^2+r^2\right)}{6 G}+\frac{M}{2}
\end{align}
and
\begin{align}
\mathcal{H}_{Pl}^{S}[\xi^z]&= \frac{k n^2}{\gamma}\left(\frac{5  \Lambda  n^2+\Lambda r^2-3}{3 G}\right)-k r n\left(\frac{5  \Lambda  n^2+\Lambda r^2-3}{3 G}\right)-2 k M n\,.
\end{align}
As before, these charges are not only divergent but also their finite parts come with the wrong factors. As  expected, adding the Euler and Pontryagin terms to Plebanski action with the coefficients as in \eqref{actionA.11} removes the divergencies and regularizes the finite part.

Note that for the Taub--NUT spacetime there is explicit dual charge contribution proportional to the Barbero-Immirzi parameter, which was absent in the case of AdS--Kerr spacetime. This contribution can be traced back to the presence of the Holst and Pontryagin terms in the Lagrangian, effectively repeating the MacDowell--Mansouri scheme of factors.

The contribution to the charges proportional to inverse $\gamma$ is subleading, and such terms are going to be present in the expressions for charges associated with corners at a finite distance. Once again we stress the regularizing effect of the topological terms; without them, the values of masses and angular momenta would not be correct (see for more details \cite{Aros:1999id, Aros:1999kt, Durka:2011yv, Durka:2012wd}).

As for the spacetimes considered before, in the AdS--Schwarzschild case the Pontryagin/Holst contribution to charges vanishes identically. In the case of AdS--Kerr the Holst/Pontryagin contributions to the charges  disappear as a result of integration over $S^2$. The situation is very different for the Taub--NUT scenario, which produces in addition the dual mass and non-zero angular momenta.


\subsection{Summary}


In this section, we have considered the asymptotic charges of some  exact asymptotically AdS solutions of Einstein equations associated with the asymptotic Killing vectors. We found that for all of them the MacDowell--Mansouri BF theory gives the finite results, while in the case of Plebanski BF theory the charges, as a rule, contain divergences and their finite part does not lead to the correct result. This means that this theory must be modified by adding terms that render the charges finite and have the right value. As it turns out the necessary counterterms are exactly the appropriate topological invariants present in the MacDowell--Mansouri BF theory action. One can speculate therefore that the proper holographic renormalization of Plebanski BF theory is the MacDowell--Mansouri BF model of gravity. To confirm this claim in the next section we will show that in the case of MM theory the asymptotic charges, the action and the symplectic structure of an arbitrary asymptotically AdS spacetimes are finite.

Finally, let us mention that the three discussed solutions of the Lagrangian \eqref{actionA.11b} have a very simple form 
\begin{align}
\mathcal{L}_{MM}&=-\frac{9}{4 \pi G\Lambda}\left(-\frac{2}{\gamma}hb+(h^2-b^2)\right)e^0\wedge e^1\wedge e^2 \wedge e^3\,,
\end{align}
where the functions $h$ and $b$ are defined for the AdS--Kerr by \eqref{h_Kerr} and \eqref{b_Kerr}, for AdS--Taub--NUT by \eqref{TNUTh} and \eqref{TNUTb}, whereas for AdS--Schwarzschild by $h=-GM/r^3$ and $b=0$. In all these cases we see that value of $\mathcal{L}_{MM}$ is finite.

At the same time, for these solutions the Plebanski Lagrangian \eqref{Pleb}  is  proportional to the volume element
\begin{align}
 32\pi G\,\mathcal{L}_{Pl}
&=\left(\frac{2}{\gamma} 4(4b-4b+\Lambda-\Lambda)+(4(2\Lambda+4h-4h)-4\Lambda)\right)e^0\wedge e^1\wedge e^2 \wedge e^3 \nonumber\\
&=
4\Lambda \, e^0\wedge e^1\wedge e^2 \wedge e^3\,,
\end{align}
which again shows the necessity of adding counterterms to this theory.


\section{General asymptotically Anti-de Sitter spacetimes}

In this section, we first briefly review the asymptotic structure of asymptotically AdS spacetime at spatial infinity following closely the classical discussion presented in \cite{Henneaux:1985tv}. Then we compute the asymptotic charges for MacDowell--Mansouri BF theory of gravity.

The AdS asymptotic metric in the coordinates $(t,r,\theta,\varphi)$ has the form
\begin{align}
 ds^2=  \bar g_{\mu \nu}dx^\mu dx^\nu =
-\left(1-\frac{r^{2} \Lambda}{3}\right) dt^2
+\frac{1}{\left(1-\frac{r^{2} \Lambda}{3}\right)} dr^2+  r^{2} d\theta^2+ r^{2} \sin^2 \theta
d\varphi^2\,.
\end{align}
The perturbations are assumed to have the following large $r$ behavior 
\begin{align}\label{asympmet}
    g_{\mu\nu}=\bar g_{\mu \nu} +  g^{(1)}_{\mu \nu} + \ldots
\end{align}
with
\begin{align}
g^{(1)}_{tt}&=\frac{h_{ t t}}{r}\,,\quad
g^{(1)}_{tr}= \frac{h_{ t r}}{r^{4}}\,,\quad 
g^{(1)}_{t\theta}= \frac{h_{ t \theta}}{r} \,,\quad
g^{(1)}_{t\varphi}= \frac{h_{ t \varphi}}{r}\,, \nonumber\\
g^{(1)}_{rr}& =\frac{h_{ r r}}{r^{5}} \,,\quad
g^{(1)}_{r\theta}= \frac{h_{ r \theta}}{r^{4}}\,,\quad
g^{(1)}_{r\varphi}=\frac{h_{ r \varphi}}{r^{4}} \,,\nonumber \\
g^{(1)}_{\theta\theta}&= \frac{h_{ \theta \theta}}{r}\,,\quad g^{(1)}_{\theta\varphi}= \frac{h_{ \theta \varphi}}{r} \,,\quad
g^{(1)}_{\varphi\varphi}= \frac{h_{ \varphi \varphi}}{r}\label{asboucond}\,,
\end{align}
where all the coefficients $h_{\mu\nu}$ are functions of $(t,\theta,\varphi)$. 

Let us consider the asymptotic corner charges, which we compute at the sphere at infinity $S=S^2_\infty$. Rewriting the general expression \eqref{H_corner_MM_explicit} in term of the metric the AdS curvature tensor $ F^{\mu\nu}{}_{\alpha\beta}=\bar{F}^{\mu\nu}{}_{\alpha\beta}+F^{(1)\mu\nu}{}_{\alpha\beta}$ leads to
\begin{align}
\mathcal{H}_{MM}^{S}[\xi] &=- \frac{3}{ 64\pi\Lambda G}\int_{S} F^{\mu\nu}{}_{\theta\varphi}\, \xi^\rho\, e^k_{\mu}\,e^l_{\nu} \left(\frac1\gamma\delta_{kl}^{ij}+\epsilon^{ij}{}_{kl}\right)\,   \omega_{\rho ij} d\theta d\varphi\,.\label{H_corner_MM_explicit_metric}
\end{align}
Let us now notice that since the background AdS part $\bar{F}$ of the AdS curvature tensor vanishes asymptotically, we can write the leading order {of perturbations} of \eqref{H_corner_MM_explicit_metric} as
\begin{align}
\mathcal{H}_{MM}^{S}[\xi] &\approx- \frac{3}{ 64\pi\Lambda G}\int_{S} F^{(1)\,\mu\nu}{}_{\theta\varphi}\, \xi^\rho\, \bar e^k_{\mu}\,\bar e^l_{\nu} \left(\frac1\gamma\delta_{kl}^{ij}+\epsilon^{ij}{}_{kl}\right)\,   \bar\omega_{\rho ij} d\theta d\varphi\,.\label{H_corner_MM_pert}
\end{align}

The AdS tetrads are defined as
\begin{align}
  \bar{e}^0=\sqrt{1- \frac{\Lambda}{3}r^{2}}\, dt,\qquad  \bar{e}^1=1/\sqrt{1- \frac{\Lambda}{3}r^{2}}\, dr,\qquad  \bar{e}^2=r\, d\theta, \qquad  \bar{e}^3=r\, \sin\theta d\varphi
\end{align}
and one derives expressions for connection $\omega^{ij}$:
\begin{align}
 \bar{\omega}^{01} &= - \frac{\Lambda}{3}r\, dt &\bar{\omega}^{23} &=- \cos\theta d\varphi\nonumber\\
\bar{\omega}^{12} &=-\sqrt{1- \frac{\Lambda}{3}r^{2}}\, d\theta &\bar{\omega}^{13} &=-\sqrt{1- \frac{\Lambda}{3}r^{2}}\, \sin\theta d\varphi\,.
\end{align}
At the same time, we have the following relevant components for the leading order asymptotic AdS curvature (all other curvature components can be found in the accompanying supplement file)
\begin{align}
F^{(1)\, t,r}{}_{\theta ,\varphi } &=\frac{3}{2r^2}  \left(\partial_\varphi h_{t\theta} - \partial_\theta h_{t\varphi} \right) \\
F^{(1)\,t,\theta }{}_{\theta ,\varphi } &= -\frac{3}{2 r^3}\,  h_{t\varphi} \\
F^{(1)\,t,\varphi }{}_{\theta ,\varphi } &= \frac{3}{2 r^3}\,  h_{t\theta} \\
F^{(1)\,r,\theta }{}_{\theta ,\varphi } &= \frac{\Lambda}{2r^2} \left(\frac{\Lambda}{9} \partial_\varphi
   h_{{rr}}-\partial_\varphi  h_{\theta \theta
   }+\partial_\theta  h_{\theta \varphi }+  \cot \theta  h_{\theta \varphi }\right) \\
F^{(1)\,r,\varphi }{}_{\theta ,\varphi } &=\frac{ \Lambda }{2r^2}\left(-\frac{\Lambda}{9} 
 \partial_\theta  h_{\text{rr}}+ \frac{1}{\sin^2\theta } (\partial_\theta h_{\varphi \varphi }-  \partial_\varphi h_{\theta \varphi })- \cot \theta  (h_{\theta \theta
   }+ \frac{1}{\sin^2\theta } h_{\varphi \varphi })\right) \\
F^{(1)\,\theta ,\varphi }{}_{\theta ,\varphi } &= \frac{ \Lambda }{2r^3}\left(-\frac{2\Lambda}{9}
    h_{\text{rr}}+  h_{\theta \theta }+ \frac{1}{\sin^2\theta} h_{\varphi \varphi }\right)\,.
\end{align}
We can further evaluate the formula \eqref{H_corner_MM_pert} to be of the form of
\begin{align}\label{Chargecompact}
\mathcal{H}_{MM}^{S}[\xi] &\approx\int_{S} \xi^\rho \,\Xi_\rho\, d\theta\, d\varphi\,,
\end{align}
where to the leading order
\begin{align}
\Xi_t&=\frac{\Lambda}{32\pi G}\sin\theta
\left(\frac{2}{9}\Lambda h_{\text{rr}}-h_{\theta\theta}-h_{\varphi\varphi}\csc^2\theta\right) +O\left(\frac1r\right)\label{Xi_t}\\
\Xi_r&=0\\
\Xi_\theta&=\frac{3}{32\pi G}\sin\theta
h_{t\theta} +O\left(\frac1r\right)\\
\Xi_\varphi&=\frac{3}{32\pi G}\sin\theta
h_{t\varphi} +O\left(\frac1r\right)\label{Xi_varphi}
\end{align}
and the metric components $h_{\mu\nu}$ are subject to the asymptotic Einstein equations (see Appendix). 
As we see the Barbero-Immirzi parameter completely disappears from these leading order asymptotic expressions. One may find this fact surprising since in the previous section we found an explicit $1/\gamma$ contribution to the charges for the AdS--Taub--NUT solution. However, this solution does not satisfy the asymptotic boundary conditions \eqref{asboucond} and therefore is not included in the general frame of this section. 

Let us note in passing that for the AdS--Kerr, as pointed out in \cite{Henneaux:1985tv}, the coordinates we used in Section 4.2 also do not have proper asymptotic behavior, which is a consequence of non-vanishing rotation at infinity. To obtain the correct value of the mass \eqref{proper_Kerr_mass} a coordinate transformation is required. Alternatively, this can be achieved by the redefinition of the Killing vector at infinity \cite{Gibbons:2004ai, Durka:2012wd}.

The asymptotic $r\rightarrow\infty$ Killing vectors are given by the following expressions that we borrowed from \cite{Henneaux:1985tv}. We took the liberty to adopt the notation used in the case of Poincar\'e Killing vectors, that could be obtained from them in the limit $\Lambda\rightarrow\infty$. In the expressions below we neglect all the terms that fall at infinity as $1/r$ or faster and we use an abbreviation $\tau \equiv \sqrt{-\Lambda/3}\, t$
\begin{itemize}
    \item Time translation
    \begin{align}
        \mathcal{E} = \frac{\partial}{\partial t}
    \end{align}
    \item Space translations
     \begin{align}
        \mathcal{P}_1 &=-\sin \tau \sin \theta \cos \varphi\,\frac{\partial}{\partial t} +r\sqrt{-\frac{\Lambda}{3}}\cos \tau \sin \theta \cos \varphi \frac{\partial}{\partial r} \nonumber\\
        &+\sqrt{-\frac{\Lambda}{3}}\cos \tau\left(\cos \theta \cos \varphi \frac{\partial}{\partial \theta}-\frac{\sin \varphi}{\sin \theta} \frac{\partial}{\partial \varphi}\right) \\
        \mathcal{P}_2 &=-\sin \tau \sin \theta \sin \varphi\,\frac{\partial}{\partial t} +r\sqrt{-\frac{\Lambda}{3}}\cos \tau \sin \theta \sin \varphi \frac{\partial}{\partial r} \nonumber\\
        &+\sqrt{-\frac{\Lambda}{3}}\cos \tau\left(\cos \theta \sin \varphi \frac{\partial}{\partial \theta}-\frac{\cos \varphi}{\sin \theta} \frac{\partial}{\partial \varphi}\right)  \\
        \mathcal{P}_3 &=- \sin \tau \cos \theta \,\frac{\partial}{\partial t} +r\sqrt{-\frac{\Lambda}{3}}\cos \tau \cos \theta  \frac{\partial}{\partial r} \nonumber\\
        &-\sqrt{-\frac{\Lambda}{3}}\cos \tau\sin \theta  \frac{\partial}{\partial \theta} 
    \end{align}
    \item Rotations
    \begin{align}
         \mathcal{M}_1 &=  \frac{\partial}{\partial \varphi} \\
         \mathcal{M}_2 &=-\sin \varphi \frac{\partial}{\partial \theta}-\operatorname{cot} \theta \cos \varphi \frac{\partial}{\partial \varphi} \\
          \mathcal{M}_3 &=\cos \varphi \frac{\partial}{\partial \theta}-\operatorname{cot} \theta \sin \varphi \frac{\partial}{\partial \varphi}
    \end{align}
    \item Boosts
     \begin{align}
        \mathcal{N}_1 &=\sqrt{-\frac{3}{\Lambda}}\cos \tau \sin \theta \cos \varphi\,\frac{\partial}{\partial t} +r\sin \tau \sin \theta \cos \varphi \frac{\partial}{\partial r} \nonumber\\
        &+\sin \tau\left(\cos \theta \cos \varphi \frac{\partial}{\partial \theta}-\frac{\sin \varphi}{\sin \theta} \frac{\partial}{\partial \varphi}\right) \\
        \mathcal{N}_2 &=\sqrt{-\frac{3}{\Lambda}}\cos \tau \sin \theta \sin \varphi\,\frac{\partial}{\partial t} +r\sin \tau \sin \theta \sin \varphi \frac{\partial}{\partial r} \nonumber\\
        &+\sin \tau\left(\cos \theta \sin \varphi \frac{\partial}{\partial \theta}-\frac{\cos \varphi}{\sin \theta} \frac{\partial}{\partial \varphi}\right)  \\
        \mathcal{N}_3 &=\sqrt{-\frac{3}{\Lambda}} \cos \tau \cos \theta \,\frac{\partial}{\partial t} +r\sin \tau \cos \theta  \frac{\partial}{\partial r} \nonumber\\
        &-\sin \tau\sin \theta  \frac{\partial}{\partial \theta} 
    \end{align}
\end{itemize}

Using the formula \eqref{Chargecompact} we can now compute all ten asymptotic charges. For example, we find the energy to be
\begin{align}
    E = \frac{\Lambda}{32\pi G}\int
\left(\frac{2}{9}\Lambda h_{\text{rr}}-h_{\theta\theta}-h_{\varphi\varphi}\csc^2\theta\right) \sin\theta d\theta d\varphi\,,
\end{align}
whereas the angular momentum is
\begin{align}
    J = \frac{3}{32\pi G}\int
h_{t\varphi}\sin\theta  d\theta d\varphi\,.
\end{align}

With the help of the formulas \eqref{Chargecompact} and \eqref{Xi_t}--\eqref{Xi_varphi} one can check that the all the remaining charges are finite, which indicates that the MacDowell--Mansouri BF theory does not require holographic renormalization. To check the finiteness of the theory we must find out further if both the Lagrangian and symplectic structure are finite as well. Let us consider the Lagrangian first. To this end, we return to the formula \eqref{actionA.11b}, which in the case of vanishing torsion can be rewritten as
\begin{align}\label{LagrMMmetric}
 32\pi G\,\mathcal{L}_{MM} 
&=-\frac{3}{2\Lambda}  \left(\frac1\gamma g_{\mu\rho}g_{\nu\sigma} -\frac1\gamma  g_{\nu\rho}g_{\mu\sigma}+\epsilon_{\mu\nu\rho\sigma}\right)\frac{1}{4}F^{(1)\mu\nu}{}_{\alpha\beta} F^{(1)\rho\sigma}{}_{\gamma\delta}\,\epsilon^{\alpha\beta\gamma\rho} \sqrt{-\bar{g}}d^4x\,.
\end{align}
This leads to the following asymptotic expansion of the Lagrangian being second perturbation order
\begin{align}
    \mathcal{L}_{MM}&=\frac{1}{r^4} \frac{2}{32 \pi  G} \sin\theta \left(\frac{\Lambda^3}{9} h_{{rr}}{}^2-2 \Lambda  h_{{rr}} \left(h_{{tt}}+\frac{\Lambda}{3}  h_{\theta \theta }\right)-9 h_{t\theta }{}^2+9 h_{\theta \theta }h_{{tt}}\right)\nonumber
 \\
 &+\frac{1}{r^4} \frac{2}{32 \pi  G } \frac{1}{\sin\theta}\left(h_{\phi \phi }\left(-\frac{2 \Lambda^2}{3} h_{{rr}}+3 \Lambda  h_{\theta \theta }+9 h_{{tt}}\right)-3 \Lambda  h_{\theta \phi }{}^2-9 h_{t\phi }{}^2\right)\nonumber\\
 &+\frac{1}{\gamma}\frac{2}{ 32 \pi G} O(1/r^5)\,.\label{asymLag}
\end{align}

It follows from \eqref{asymLag} that the action is finite (assuming appropriate boundary conditions at the horizons in the bulk, if any are present). As a consequence, the symplectic structure that can be derived from the  action is also finite. This concludes our proof that the MacDowell--Mansouri BF action for gravity is well suited for describing physics in asymptotically AdS spacetime.


\section{Conclusions}

In this paper, we considered two constrained BF theories of gravity, one formulated by Plebanski and another based on MacDowell--Mansouri theory. Both give the same bulk Einstein equations with the negative cosmological constant, but since their actions differ by the presence of topological invariants in the second case, the expressions for charges associated with corners, at infinity, and in the final distance are different. We show that as a rule the asymptotic charges are divergent for Plebanski theory with the finite part also giving incorrect results. On the other hand for the MacDowell--Mansouri theory, all the charges are finite and have the {correct}  value. Moreover, the action of this theory and the symplectic structure are both finite. This indicates that to properly describe gravity in asymptotically AdS spacetime one should use the MacDowell--Mansouri theory and not the Plebanski one.


\appendix
\section{Asymptotic Einstein equations}\label{appendix}

The Einstein equations can be easily produced from the evaluation of the contacted AdS curvature, namely
\begin{align}
F^{\lambda}{}_{\mu \lambda \nu } &=R^{\lambda}{}_{\mu \lambda \nu }-\frac{\Lambda }{3}(g^{\lambda}{}_{\lambda}g_{\mu \nu}-g^{\lambda}{}_{\nu}g_{\mu\lambda})=R_{\mu \nu }-\frac{\Lambda }{3}(4 g_{\mu \nu }-g_{\mu \nu }) \nonumber\\
&=R_{\mu \nu }-\frac{3\Lambda }{3}g_{\mu \nu }=G_{\mu v}\,,
\end{align}
as the Einstein tensor is then $G_{\mu v}=R_{\mu v}-\frac{1}{2}g_{\mu v}R+\Lambda
g_{\mu v}=R_{\mu v}-\Lambda g_{\mu v}$. 

The Einstein equations decompose accordingly to $0=G_{\mu\nu}=\bar{G}_{\mu\nu} +G^{(1)}_{\mu\nu}$, where AdS part is identically vanishing. Their leading order in the  powers of $r$  are:
\begin{itemize}
\item $G^{(1)}_{tt}$ of order $1/r$
    \begin{align*}
     0=   \frac{\Lambda^2}{3} h_{rr}-3
   h_{tt}-\Lambda  \left(h_{\theta
   \theta}+\csc ^2\theta h_{\varphi \varphi}\right)
    \end{align*}
    
 \item $G^{(1)}_{tr}$ of order $1/r^4$
    \begin{align*}
     0=  \frac{2 \Lambda}{9} \partial_t h_{rr}- \partial_t h_{\theta\theta}+
   \partial_\theta h_{t\theta}+ \cot \theta h_{t\theta}- \csc ^2\theta \partial_t h_{\varphi \varphi}+ \csc ^2\theta \partial_\varphi h_{t\varphi}
    \end{align*}
    
 \item $G^{(1)}_{t\theta }$ of order $1/r^3$
    \begin{align*}
       0&= \frac{2\Lambda}{3}   \partial_t h_{r\theta }+\frac{\Lambda}{3} 
  \partial_t \partial_\theta  h_{{rr}}-2 h_{t\theta }+\frac{2}{3} \Lambda  \partial_\theta  h_{{tr}}+\cot \theta \partial_t h_{\theta \theta }\\
   &+ \csc ^2\theta\left( \partial_t \partial_\varphi h_{\theta \varphi }-  \partial_t \partial_\theta h_{\varphi
   \varphi }+\cot \theta  \partial_t h_{\varphi \varphi
   }-  \partial^2_\varphi
   h_{t\theta }+  \partial_\theta \partial_\varphi h_{t\varphi }\right)  
    \end{align*}
    
\item $G^{(1)}_{t\varphi }$ of order $1/r^3$
    \begin{align*}
        0&= \Lambda \partial_t \partial_\varphi h_{rr}+2 \Lambda \partial_t h_{r\varphi
   }-3 \partial_t \partial_\varphi h_{\theta \theta }+3 \partial_t \partial_\theta h_{\theta \varphi
   }+3 \cot \theta \partial_t h_{\theta \varphi }\\
   &+3 \partial_\theta \partial_\varphi
   h_{t\theta}-3 \cot \theta \partial_\varphi h_{t\theta}+2 \Lambda  \partial_\varphi h_{tr}-6
   h_{t\varphi}-3 \partial^2_\theta h_{t\varphi}+3 \cot (\theta
   )  \partial_\theta  h_{t\varphi}
    \end{align*}
    
\item $G^{(1)}_{rr}$ of order $1/r^5$
    \begin{align*}
        0= \frac{\Lambda^2}{3} h_{rr}-3
   h_{tt}-\Lambda  \left(h_{\theta
   \theta }+\csc ^2\theta h_{\varphi \varphi }\right)
    \end{align*}
    
\item $ G^{(1)}_{r\theta }$ of order $1/r^4$ 
\begin{align*}
0&=\frac{2\Lambda ^2}{9}   \partial_\theta  h_{rr}+3 \partial_t h_{t\theta}-3  \partial_\theta h_{tt}+ \Lambda  \cot \theta
   h_{\theta \theta } + \Lambda  \cot (\theta
   ) \csc ^2\theta h_{\varphi \varphi }\\
  & + \Lambda  \csc ^2\theta \partial_\varphi h_{\theta \varphi }- \Lambda  \csc ^2\theta  \partial_\theta  h_{\varphi \varphi}
\end{align*}

\item $ G^{(1)}_{r\varphi}$ of order $1/r^4$ 
\begin{align*}
0&=2 \Lambda ^2 \partial_\varphi h_{rr}-9 \Lambda  \partial_\varphi h_{\theta \theta
   }+9 \Lambda   \partial_\theta  h_{\theta \varphi }+9 \Lambda  \cot
   \theta h_{\theta \varphi }-27 \partial_\varphi h_{tt}+27
   \partial_t h_{t\varphi}
\end{align*}

\item $ G^{(1)}_{\theta\theta}$ of order $1/r$  
\begin{align*}
0&=\frac{\Lambda^2}{3} h_{rr}-3
   h_{tt}-\Lambda  \left(h_{\theta
   \theta }+\csc ^2\theta h_{\varphi \varphi }\right)
\end{align*}

\item $ G^{(1)}_{\theta\varphi}$ of order $1/r^3$  
\begin{align*}
0&=2 \Lambda ^2 \partial_\varphi h_{r\theta}+\Lambda ^2 \partial_\theta \partial_\varphi
   h_{rr} +2 \Lambda ^2  \partial_\theta h_{r\varphi} -18 \Lambda  h_{\theta \varphi }\\
   &-9 \partial^2_t h_{\theta \varphi
   }+9 \partial_t \partial_\varphi h_{t\theta}-9 \partial_\theta \partial_\varphi
   h_{tt}+9\partial_t \partial_\theta  h_{t\varphi}\\
   &-\Lambda ^2 \cot \theta \partial_\varphi h_{rr}-4 \Lambda ^2 \cot \theta
   h_{r\varphi}+9 \cot \theta \partial_\varphi h_{tt}-18 \cot \theta \partial_t h_{t\varphi
   }
\end{align*}

\item $G^{(1)}_{\varphi\varphi}$ of order $1/r$  
\begin{align*}
0&=\frac{\Lambda^2}{3} h_{rr}-3
   h_{tt}-\Lambda  \left(h_{\theta
   \theta }+\csc ^2\theta h_{\varphi \varphi }\right) 
\end{align*}
\end{itemize}
Note the repeated equation in the diagonal part, which validity can be tested for instance by using the AdS-Schwarzschild perturbation scheme with
\begin{align}
    h_{tt}=2 G M\,,\qquad h_{rr}=2 G M \frac{9}{\Lambda^2}\,.
\end{align}


\section*{Acknowledgment}

For JKG, this work was supported by funds provided by the National Science Center, project number 2019/33/B/ST2/00050.



\end{document}